# Moderate hybridization effects revealed by ARPES in heavy-fermion $Ce_2IrIn_8$

H.-J. Liu,[1,2] Y.-J. Xu,[1,2] Y.-G. Zhong,[1,2] J.-Y. Guan,[1,2] L.-Y. Kong,[1,2] J.-Z. Ma,[3] Y.-B. Huang,[4] Q.-Y. Chen,[5] G.-F. Chen,[1,2] M. Shi,[3] Y.-F. Yang,[1,2] H. Ding,[1,2,6†]

[1]*Beijing National Laboratory for Condensed Matter Physics and Institute of Physics, Chinese Academy of Sciences, Beijing 100190, China*
[2]*School of Physics, University of Chinese Academy of Sciences, Beijing 100190, China*
[3]*Paul Scherrer Institute, Swiss Light Source, CH-5232 Villigen PSI, Switzerland*
[4]*Shanghai Synchrotron Radiation Facility, Shanghai Institute of Applied Physics, Chinese Academy of Sciences, Shanghai 201204, China*
[5]*Science and Technology on Surface Physics and Chemistry Laboratory, Mianyang 621908, China*
[6]*Songshan Lake Materials Laboratory, Dongguan, Guangdong 523808, China*

† Corresponding author: dingh@iphy.ac.cn

## ABSTRACT

We utilized high-resolution resonant angle-resolved photoemission spectroscopy (ARPES) to study the band structure and hybridization effect of the heavy-fermion compound $Ce_2IrIn_8$. We observe a nearly flat band at the binding energy of 7 meV below the coherent temperature $T_{coh} \sim 40$ K, which characterizes the electrical resistance maximum indicating the onset temperature of hybridization. However, the Fermi vector $k_F$ and the Fermi surface (FS) volume have little change around $T_{coh}$, challenging the widely believed evolution from a high-temperature small FS to a low-temperature large FS. Our experimental results of the band structure fit well with the density functional theory plus dynamic mean-field theory (DFT+DMFT) calculations.

## INTRODUCTION:

Heavy-fermion compounds, first discovered in $CeAl_3$ in 1975 [1], are some of the most exotic materials in condensed matter physics. The name originates from the largely enhanced effective mass of the heavy quasi-particles, which can be 2 or 3 orders of magnitude higher than that in a normal metal [2]. These compounds usually contain some of Ce, Sm, Yb, U, Pr, Pu, Np elements, which possess an unfilled 4f or 5f shell. It is widely believed that 4f/5f electrons are local moments at high temperatures and become itinerant after hybridized with the conduction electrons at low temperatures. Varieties of phenomena, e.g., antiferromagnetism [3], ferromagnetism [4], superconductivity [5], quantum critical point [6], quadrupole order [7], hidden order [8-9], topological property [10], have been discovered in heavy-fermion compounds. Central to understanding these exotic phenomena is the interplay of itinerancy and localization. However, the low energy scales (critical temperature, hybridization gap, superconducting gap) in heavy-fermion systems have brought major challenges to many experimental techniques.

$Ce_mT_nIn_{3m+2n}$ (m = 1, 2; n = 1, 2 and T: Co, Rh, Ir, Pd, Pt) family is a good platform of heavy-fermion materials for studying the interplay between c-f hybridization, magnetism, superconductivity, quantum criticality, and etc. $Ce_mT_nIn_{3m+2n}$ crystallizes with a tetragonal unit cell that can be viewed as m-layers of $CeIn_3$ unit stacked sequentially with intervening n-layers of $TIn_2$ along the c-axis. Among them, the spin-glass state observed in $Ce_2IrIn_8$ indicates partially delocalized Ce 4f electron [11]. The magnetism in $Ce_2IrIn_8$ depends on Ce-Ir hybridization and local Ce environment. The small but finite onset temperature for spin freezing rules out the quantum critical point (QCP) scenario in $Ce_2IrIn_8$ [11]. High Sommerfeld coefficient ($\gamma \sim 700$ mJ/mol·$K^2$) [12] and the absence of long-range magnetic order indicate an itinerant behavior of Ce 4f electron. μSR Knight-shift experiments [13-14] observed a 'Knight-shift anomaly' in which the Knight shift constant K no longer scales linearly with the susceptibility below a characteristic temperature $T_{coh}$, in agreement with the "two-fluid" model of heavy-fermion formation [15]. Resistivity measurements showed a broad maximum near 40 - 50 K [13,16], manifesting the development of a coherent state. ARPES is a powerful tool to directly measure the electronic structure and Fermi surface of solid state materials. Especially, resonant ARPES with photon energies near 120 eV has been proved to be effective in tracing subtle changes of the 4f electrons in the Ce-based heavy-fermion compounds [17-19]. However, systematic ARPES research on the electronic structure of $Ce_2IrIn_8$ is still lacking to reveal the hybridization process in its ground state.

In this paper, we performed a systematic electronic structure study on $Ce_2IrIn_8$ using ARPES measurements and DFT+DMFT calculations to investigate the localized/itinerant nature of the f electrons in its ground state. We found a nearly flat band at the Γ point at low temperatures. Bands away from the Γ point have no flat character, indicating the Ce 4f-electron contribution appears mainly around the Γ point, which is similar with other cerium based heavy-fermion compounds [17-19]. We found that $Ce_2IrIn_8$ has a similar band structure with the $CeMIn_5$ (M = Co, Rh, Ir) compounds near the Fermi level: hole pockets around the Γ point and electron pockets around the M point. Neither in our experimental nor theoretical explorations, we could find large change of the Fermi surfaces at very low temperature, which is different from the widely believed evolution from small Fermi surfaces at high temperatures to large Fermi surfaces at low temperatures.

### EXPERIMENTAL AND COMPUTATIONAL DETAILS:

High-quality single crystals of $Ce_2IrIn_8$ were grown by flux method. Electrical resistivity was measured with the four-probe method. ARPES data in Fig. 1 were obtained at the SIS beamline of Swiss Light Source using a SCIENTA R4000 analyzer. Samples were cleaved *in situ* along the (001) plane at T = 15 K, the vacuum was kept below $5 \times 10^{-11}$ mbar. Other ARPES data were obtained at the "Dreamline" beamline of Shanghai Synchrotron Radiation Facility (SSRF). Samples were cleaved *in situ* at T = 10 K, and the vacuum was kept below $1 \times 10^{-10}$ mbar. The overall energy resolution was better than 18 meV.

We used the DFT+DMFT method [20-22] with the continuous time quantum Monte Carlo (CTQMC) [23] and one-crossing approximation (OCA) [24] as the impurity solver for the first-principles calculations of this compound. The lattice is represented using the full potential linear augmented plane wave method with spin orbital coupling as implemented in WIEN2k package [25] and the Perdew-Burke-Ernzehof generalized gradient approximation (PBE-GGA) is used [26]. We chose the Coulomb interaction U = 6.0 eV to yield the correct coherence temperature and obtain the self-energy, the density of states and the spectral function in real frequency with analytical continuation in the CTQMC calculations, which were then compared to the OCA real frequency calculations for consistency.

### RESULTS AND DISCUSSIONS:

The electrical resistivity of $Ce_2IrIn_8$ shown in Fig. 1(a) has a broad maximum at T ~ 40 K, which is the coherent temperature of $Ce_2IrIn_8$. To investigate the basic electronic properties of $Ce_2IrIn_8$, we measured the Fermi surface and band structure of $Ce_2IrIn_8$ at T = 16 K. All the ARPES data in Fig. 1 were obtained with an off-resonant photon energy of 115 eV, which has a small cross-section for Ce 4f electron and can reveal the non-4f band more clearly. The Fermi surface in the Γ-X-M plane, shown in Fig. 1(b), is composed of 5 electron pockets. Around the Γ point, there is an electron pocket α with the four-corner star-like shape. The bottom of this electron pocket α is around 0.29 eV below the Fermi level, which is consistent with the Ce $4f_{7/2}$ final state [27,28]. Around the M point, there are three electron pockets γ,δ,ε, with their bottoms around 0.51 eV, 0.76 eV, 1.4 eV below the Fermi level with a radius of 0.29 $A^{-1}$, 0.36 $A^{-1}$, 0.5 $A^{-1}$, respectively. Around the X point, there is an elongated electron pocket β with its long axis along the Γ-X direction. Because the slit of the analyser in the SIS beamline is parallel to the ground, so linear horizon (LH) polarization can be used to detect even parity orbitals and z components, while linear vertical (LV) polarization is used for detecting pure odd parity orbitals. Comparing Figs. 1(c) and 1(d), the electron pocket α near the Γ point is more obvious with LH polarization along both Γ-X and Γ-M directions, implying the electron pocket α should not be $d_{xy}$ or $d_{x^2-y^2}$ orbital. Due to the $d_{xz}$ and $d_{yz}$ orbitals are purely even/odd with respect to the ac mirror plane, and we cannot detect the electron pocket with LV polarization, indicating that the electron pocket α near the Γ point should be the $d_{z^2}$ orbital. For other bands near the Fermi level, although there is intensity difference, they can be detected by both LH and LV polarizations, indicating they are not pure orbitals.

Resonant ARPES measurements were performed at Ce 4d-4f transition [28] to enhance the Ce 4f cross-section in $Ce_2IrIn_8$. We choose hν = 121 eV as the resonant photon energy, and hν = 115 eV as the off-resonant photon energy to minimize the cross-section and kz difference between on/off-resonant photon energy. Figs. 2(a) and 2(b) show a photoemission intensity plot taken with off-resonant (hν = 115 eV) and on-resonant (hν = 121 eV) photons at T = 15 K along the Γ-M direction with a circular polarization, respectively. Compared with the on-resonant spectra, the conduction bands dominate the off-resonant spectra. The Ce 4f component is strongly enhanced in the on-resonant data, as also shown in the angle-integrated energy distribution curves (EDCs) in Fig. 2(c). Two nearly flat features reside in 0.3 eV and 2.5 eV below the Fermi level can be observed in the on-resonant data. Two peaks highlighted in Fig. 2(c) correspond to the Ce $4f^1_{7/2}$ and Ce $4f^0$ final states [28-30].

Furthermore, fine features near the Fermi level are shown in Figs. 2(d) and 2(e) with a smaller energy window. Fig. 2(g) shows the EDCs of Figs. 2(d) and 2(e) at the Γ point. Compared with Fig. 2(d), we observed an obvious spectral weight enhancement in the on-resonant data around the Γ point near the Fermi level in Fig. 2(e). This enhanced intensity at the Γ point near the Fermi level is actually the tail of Kondo resonance [31]. From the on/off-resonant comparison, we can find that resonant ARPES not only enhances the Ce 4f cross-section due to the 4d-4f transition, but also enhances the background making the other bands fuzzy.

To investigate the $k_F$ difference due to the f-electron contribution in resonant spectrum, momentum distribution curves (MDCs) at $E_F$ are displayed in Fig. 2(f). We draw three vertical dashed lines to mark the $k_F$ positions away from the Γ point in Fig. 2(f). As shown in Fig. 2(f), the on/off resonant MDCs have no visible difference at these marked $k_F$ positions. However, on-resonant ARPES suppresses the intensity at $k_F$ = ± 0.65 $A^{-1}$, but doesn't change the $k_F$ value, suggesting that there is no 4f contribution involved in this electron pocket centered at the M point, or the hybridization happens far above the Fermi level. The only intensity enhancement in the resonant MDC at the Fermi level is around the Γ point. There is an emerging intensity near the Γ point compared with the off-resonant (hν = 115 eV) MDC, which indicates that the Ce 4f feature near $E_F$ mainly concentrate around the Γ point. Furthermore, $k_F$ shrinks in the resonant MDC, which also indicates that the f ingredient appears mostly near the Γ point.

In order to trace the hybridization process in $Ce_2IrIn_8$, we performed a temperature dependent experiment with the on-resonant photon energy (hν = 121 eV) along the Γ-M direction. At the lowest temperature (T = 10 K), we observed a nearly flat band near the Γ point at the binding energy about 7 meV. The intensity of this flat band gradually decreases with increasing temperature and nearly disappears above the coherent temperature $T_{coh}$ = 40 K. The EDCs show distinct change from a low-temperature peak near the Fermi level to a high-temperature Fermi-Dirac cutoff line shape near the Fermi level in Fig. 3(e), which indicates the hybridization process happens at low temperatures. Integrating EDCs over the energy range of ($E_F$ - 100 meV, $E_F$ + 10 meV), we obtain the T-dependent f-electron spectral weight, as shown in the inset of Fig. 3(e), which displays an obvious increase upon lowering T. From our data, we get the similar enhancement value (~ 20%) as the weakly hybridized heavy-fermion compounds $CeRhIn_5$ and $CeCoIn_5$ [18]. Furthermore, T-dependent MDCs at the Fermi level can in principle reflect the change of $k_F$ around the coherent temperature. However, three vertical dashed lines in Fig. 3(f) line up the MDC peaks at different temperatures nicely, indicating very small difference in $k_F$ when temperature changes across $T_{coh}$, which is quite unexpected and contradict with the small-to-large Fermi surface change scenario expected in these heavy-fermion compounds. However, as we will discuss below, this might occur if there exist several unconnected Fermi surfaces. Similar with our result, there is also no obvious change [18] in $k_F$ in $CeRhIn_5$, even there is a flat band with a sharp peak near $E_F$.

To understand these results, we carried out DFT+DMFT calculations with experimental lattice parameters for $Ce_2IrIn_8$. Figs. 4(a) and 4(b) plot the temperature evolution of the Ce-4f density of states. We see a sharp quasiparticle resonance developing near the Fermi energy at T = 10 K and a rapid increase of the quasiparticle peak when lowering the temperature. The onset temperature of the increase corresponds to the maximum of the temperature derivative of the imaginary part of the self-energy. For U = 6 eV, this occurs at about 30 K, in a rough agreement with the coherence temperature estimated from experiment, which justifies our choice of parameters. The band structure at 10 K is plotted in Fig. 4(e) and exhibits evident hybridization feature near the Fermi energy, which is absent at 200 K. Several features in the corresponding Fermi surfaces are consistent with experiment. Along the X-M direction are four electron bands, two of which are close and difficult to distinguish in ARPES. The bottoms of these bands locate at 0.8 eV and 1.4 eV at the M point, respectively, in good agreement with experiment. These bands show negligible f-character and hence remain unchanged at all temperatures. By contrast, the major f-electron bands at 10 K locate around the Γ point and along the Γ-X direction, forming a complicated pattern of Fermi surfaces shown in Figs. 4(d) and 4(f), which evolves with lowering temperature due to the local-to-itinerant transition. For example, around the Γ point there exists an electronic pocket, which increases slightly with lowering temperature. This difference agrees with the experimental observation. We also note that the overall change is small compared with the usual expectation of small-to-large Fermi surface change. But this should not be surprised, if one notices the existence of several electron and hole Fermi surfaces. While each of them varies slightly, a sum of them all would yield the required volume change by the Luttinger theorem. In our ARPES data at hv = 121 eV, each momentum pixel is $0.0095 A^{-1}$, if we assume each of the three electron pockets γ,δ,ε, centered at M point expands two momentum pixels, we estimate the total Fermi volume expansion to be about 0.078 electrons in the localized-to-itinerant transition. If we add the contributions from the α, β electron pockets, we may get similar value $0.2\pm0.05$ in $CeCoIn_5$. [17]

## CONCLUSIONS

To summarize, we performed high-resolution on-resonant temperature dependent study on the heavy-fermion compound $Ce_2IrIn_8$ across the coherent temperature. We found a nearly flat band near the Fermi level around the Γ point showing no big difference in $k_F$, which seems to challenge the small-to-large Fermi surface scenario in the heavy-fermion compounds. In our case, f-electron orbitals mainly exist near the Γ point, but without large changes for $k_F$ values and Fermi surface volume. This may due to multiple Fermi surfaces in this compound. Each single Fermi surface volume varies slightly with temperature, and the total effect of multiple Fermi surfaces yields a moderate volume change contributed from Ce 4f electrons.

**Acknowledgements:**

We thank G. Phan and D. Chen for modification of the manuscript. This work at IOP is supported by the grants from the Ministry of Science and Technology of China (2016YFA0401000, 2015CB921300, 2016YFA0300303, 2017YFA0303103), the Natural Science Foundation of China (11674371, 11774401, 11874330), the Chinese Academy of Sciences (XDB07000000), and the Beijing Municipal Science and Technology Commission (Z171100002017018). Y.-B. Huang acknowledges supports by the Ministry of Science and Technology of China (2016YFA0401002) and the CAS Pioneer "Hundred Talents Program" (type C). M. Shi is supported by the Sino-Swiss Science and Technology Cooperation (Grant No. IZLCZ2-170075) and the Swiss National Science Foundation (No. 200021-159678)

H. Ding designed the experiments and supervised the project. H.-J. Liu carried out the ARPES experiments with contributions from Y.-G. Zhong, J.-Y. Guan, L.-Y. Kong, J.-Z. Ma, Y.-B. Huang. Y.-J. Xu and Y.-F. Yang did the calculations. G.-F. Chen synthesized the single crystals. H.-J. Liu and H. Ding performed the data analysis, figure development, and wrote the paper with contributions from Q.-Y. Chen, Y.-F. Yang, M. Shi. All authors discussed the results and interpretation.

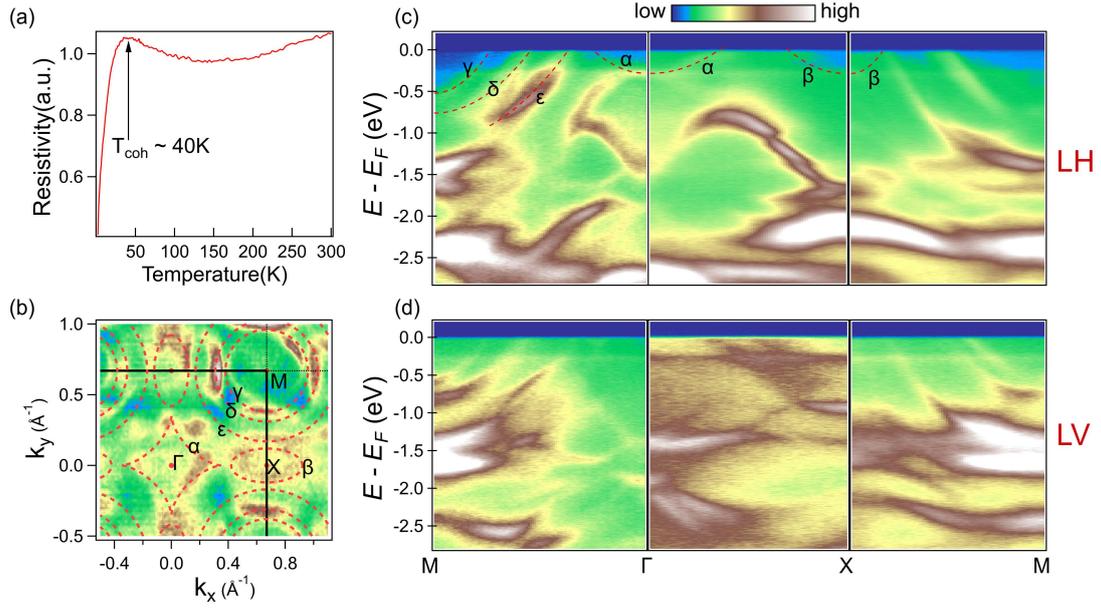

FIG. 1. Electrical resistivity, Fermi surface and polarization dependent band structure along the high symmetry axes. (a) Electrical resistivity of $Ce_2IrIn_8$. (b) ARPES intensity plot integrated over a window of ($E_F$ - 10 meV, $E_F$ + 10 meV) with circular polarization photons, representing Fermi surfaces of $Ce_2IrIn_8$. (c) Band dispersions along M-Γ-X-M momentum path with LH polarization. The red dashed pockets in (b) and the red dashed lines in (c) indicate the Fermi pockets and corresponding band dispersions, respectively. (d) same with (c) but with LV polarization. All the ARPES data were obtained with an off-resonant photon energy hν = 115 eV, at T = 16 K.

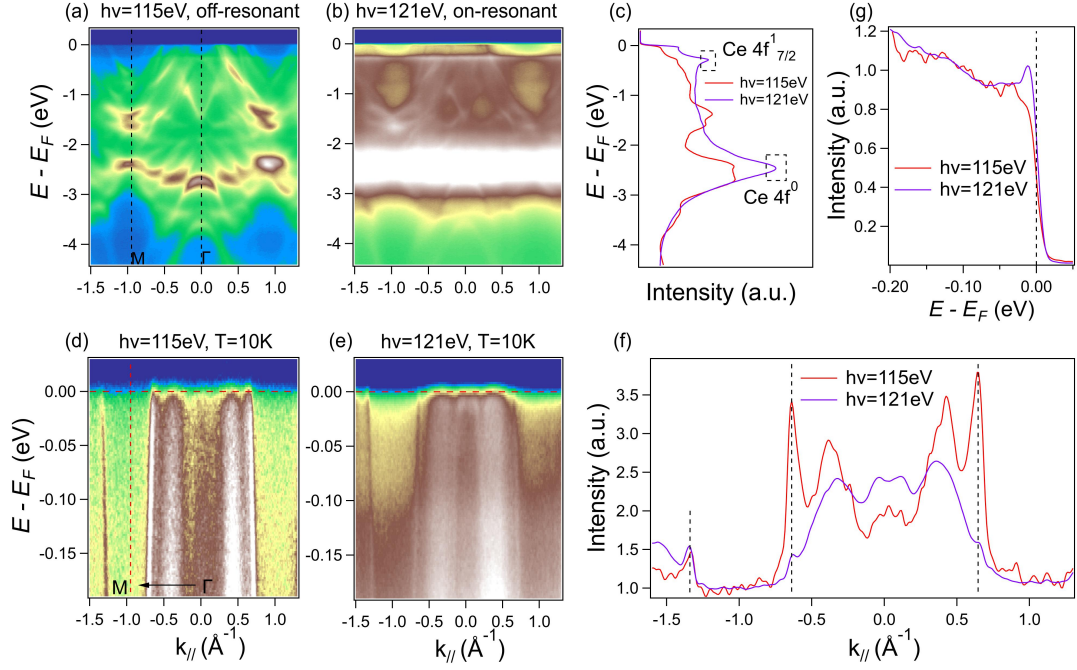

FIG. 2. On/off-resonant ARPES data of $Ce_2IrIn_8$ at low temperatures. (a) Off-resonant (hv = 115 eV) intensity plot along the Γ-M direction at T = 15 K. (b) Same with (a) but with the on-resonant (hv = 121 eV) photon energy. (c) The angle-integrated EDCs of (a) and (b). (d) Off-resonant intensity plot taken along the Γ-M direction with hv = 115 eV, T = 10 K. The vertical dashed red line indicates the position of M. (e) On-resonant intensity plot with hv = 121 eV, T = 10 K. (f) On/off resonant MDCs at the Fermi level. (g) On/off resonant EDCs at the Γ point.

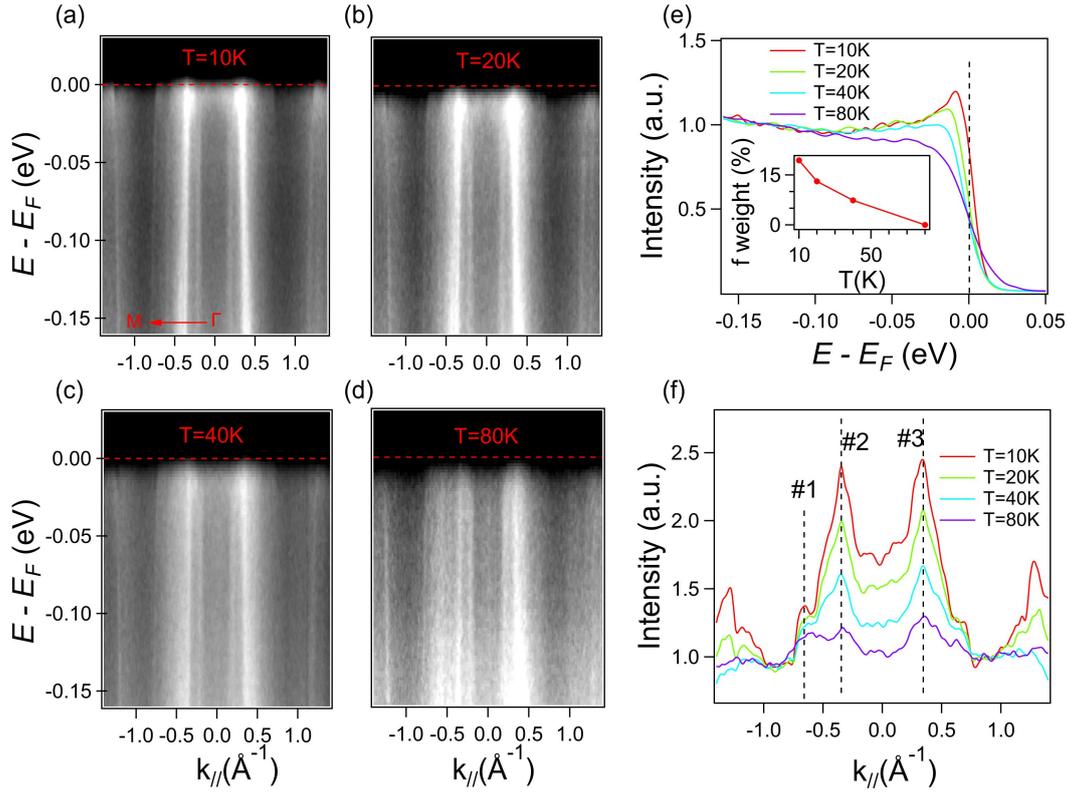

FIG. 3. Temperature dependent evolution of band dispersion along the Γ-M direction. Intensity plots taken with hv = 121 eV at (a) T = 10 K, (b) T = 20 K, (c) T = 40 K, (d) T = 80 K. (e) EDC curves taken at the Γ point at different temperatures, inset is f-electron weight derived from EDCs. (f) MDCs taken at the Fermi level at different temperatures, the vertical dashed lines mark the $k_F$ positions.

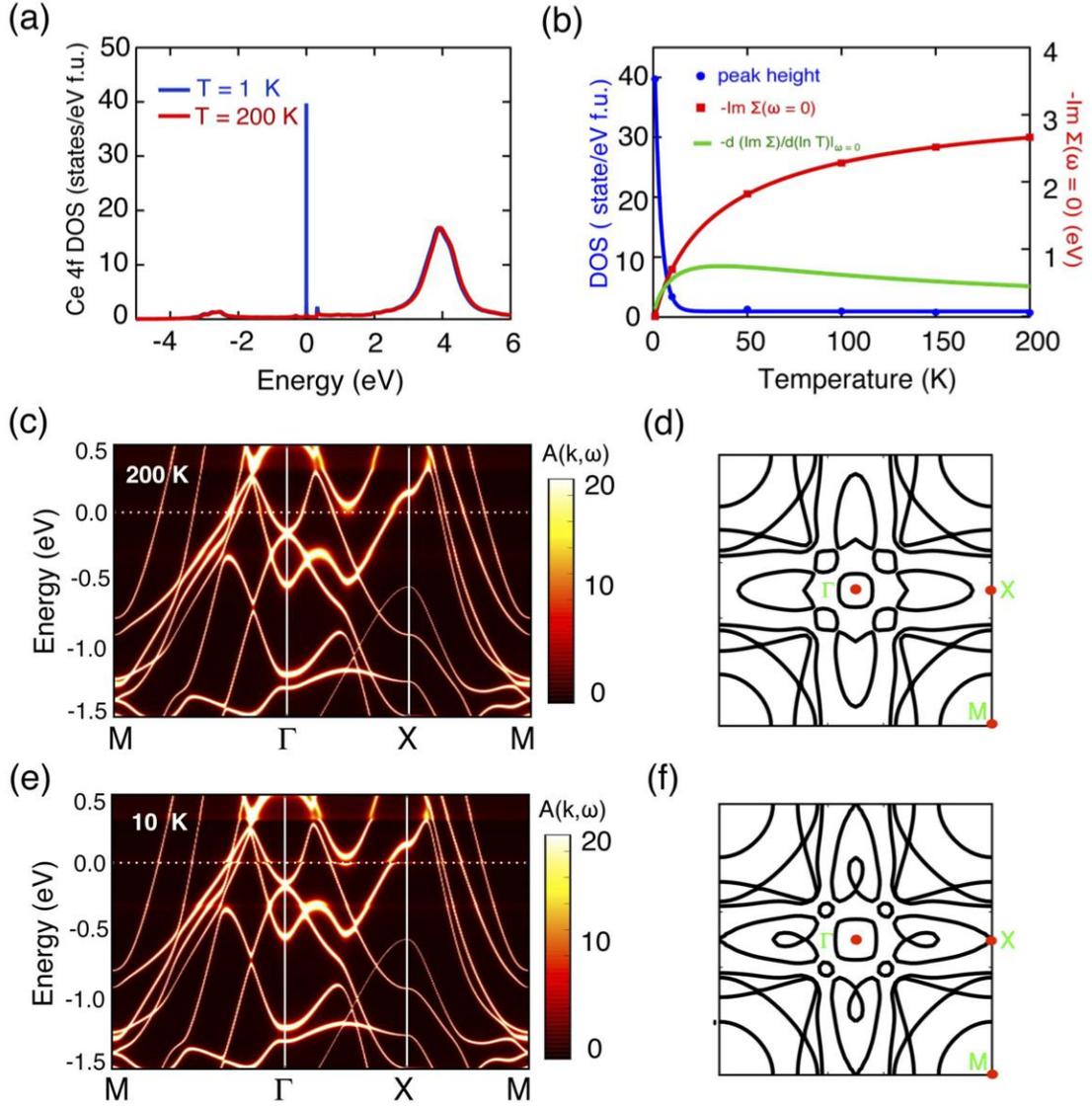

FIG. 4. Temperature dependent evolution of density of states, spectral function and Fermi surfaces. (a) The density of states of Ce-4f electrons at high temperature of T = 200 K and low temperature at T = 1 K. (b) Temperature evolution of the height of quasiparticle peak, the imaginary part of Ce-4f self-energy at the Fermi energy and its temperature derivative. (c) and (e) are the momentum-resolved spectral functions at T = 200 K and T = 10 K, respectively. (d) and (f) are the corresponding Fermi surfaces at T = 200 K and T = 10 K, respectively.